\documentclass[9pt,twocolumn,twoside]{opticajnl}

\journal{opticajournal} 

\setboolean{shortarticle}{true}

\title{Subdiffusive dynamics in photonic random  walks probed with classical light}

\author[1,2,*]{Stefano Longhi}

\affil[1]{Dipartimento di Fisica, Politecnico di Milano, Piazza L. da Vinci 32, I-20133 Milano, Italy}
\affil[2]{IFISC (UIB-CSIC), Instituto de Fisica Interdisciplinar y Sistemas Complejos - Palma de Mallorca, Spain}

\affil[*]{stefano.longhi@polimi.it}

\begin{abstract}
The random walk of photons in a tight-binding lattice is known to exhibit diffusive motion similar to classical random walks under decoherence, clearly illustrating the quantum-to-classical transition. In this study, we reveal that  the random walk of intense classical light under dephasing dynamics can disentangle quantum and ensemble averaging, making it possible to observe a subdiffusive walker dynamics, i.e. a behavior very distinct from both a classical and a quantum walker. 
 These findings are demonstrated through proposing photonic random walks in synthetic temporal lattices, based on pulse dynamics in coupled fiber loops.
\end{abstract}

\setboolean{displaycopyright}{false} 

\begin{document}

\maketitle

{\em Introduction.}  
The quantum-to-classical transition and decoherence are critical concepts in understanding the behavior of quantum systems as they interact with their environment \cite{R1}. Quantum random walks (QRWs), the quantum analogs of classical random walks, have emerged in the past few decades as a significant area of research (see e.g. \cite{R2,R3,R4,R5} and references therein).  Besides offering a basic framework for developing effective quantum algorithms and simulating complex phenomena \cite{R2,R3,R4}, QRWs serve as a valuable model for studying decoherence effects and quantum-to-classical transitions, providing insights into how quantum coherence is lost and classical behavior emerges \cite{R5,R6,R7,R8,R9,R10,R11,R12,R13,R14,R15,R16,R17,R18,R19,R20}. To this regard, photonics has offered a fascinating and experimentally-accessible platform to implement quantum walks and to test the quantum to classical transition \cite{R15,R16,R21,R22,R23,R23b,R23c,R23d,R24}, enabling precise control over the quantum states and the introduction of controllable decoherence via dynamical stochastic phases \cite{R15,R16}. In a QRW on the line, the quantum-to-classical transition is observed through the change in the spreading pattern of the probability distribution of the photon position. This transition is characterized by a shift from ballistic spreading, typical of a coherent QRW, to diffusive spreading, which is indicative of a classical random walk (CRW) \cite{R5}. In the strong dephasing regime the photon undergoes an incoherent hopping motion on the lattice like a classical walker, resulting in a characteristic diffusive motion. The probability distribution of the photon in a QRW with dephasing arises from a combination of quantum randomness and ensemble averaging over stochastic phase realizations, which is fully captured by the density matrix of the photon field. This dual stochastic process is central to understanding classicalization of the walker's behavior. When the walker is realized by an intense classical light field carrying a large number of photons, the measurement procedure can disentangle quantum randomness from ensemble randomness, leading to a completely different walker behavior. Namely, a subdiffusive behavior can emerge, which is very distinct than one expected for either a classical or a quantum walker. Such a peculiar subdiffusive behavior was predicted in some earlier numerical studies in certain disordered systems \cite{R25,R26,R27},  however its relevance in quantum walks and the possibility  of disentangling quantum and ensemble randomness in accessible experimental platforms remain largely unexplored.\\  
 In this Letter, we explore subdiffusive motion arising from the disentanglement of quantum and ensemble randomness in a photonic random walk on a one-dimensional lattice under strong dephasing dynamics. We demonstrate that subdiffusive behavior can be understood through an effective two-dimensional random walk with defects in occupation probability correlations, offering analytical insights into the problem. Additionally, we propose an experimentally accessible setup based on optical pulse dynamics in photonic temporal mesh lattices \cite{R28,R29,R30,R31}, where the use of intense classical light fields should provide a straightforward method to disentangle quantum and ensemble randomness.\\
{\it Classicalization of photonic quantum walks by dephasing.}  Let us consider continuous-time quantum walks of photons on a one-dimensional tight-binding lattice, such as in coupled optical waveguides or resonators \cite{R31b} [Fig.1(a)] or in synthetic lattices in time or frequency dimensions. The coherent dynamics  of the photon field is described by the tight-binding Hamiltonian
\begin{equation}
\hat{H}=\sum_n J  \left( \hat{a}^{\dag}_n \hat{a}_{n+1}+{\rm H.c.} \right)
\end{equation}
 \begin{figure}[h]
 \centering
   \includegraphics[width=0.42\textwidth]{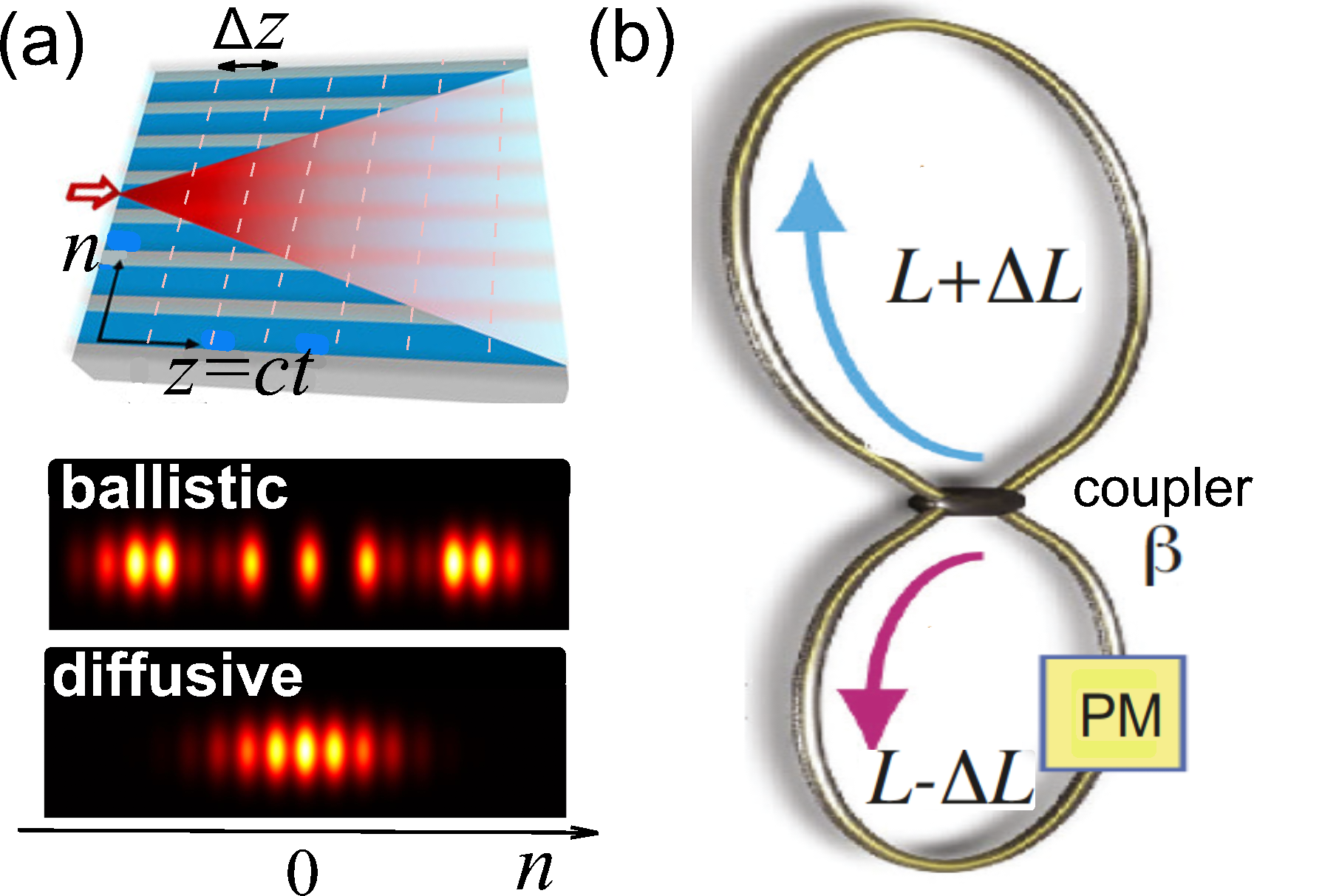}
   \caption{ \small (a) Schematic of continuous-time photonic quantum walk on a one-dimensional waveguide lattice. A single photon is injected into site $n=0$, as indicated by the red arrow. Under coherent dynamics, light spreads ballistically  
   along the lattice. Classicalization of the quantum walk, leading to diffusive spreading, is obtained by applying stochastic phases after each propagation step $\Delta z= c \Delta t$ (dashed lines in the plot).  For illustrative purpose, the bottom panels show the averaged light intensity distributions after the propagation distance $z=ct=3.5 c/J$, corresponding to ballistic and diffusive spreading dynamics, observed by classical light excitation. (b) Photonic random walk in a synthetic temporal lattice realized by optical pulse propagation in two coupled fiber loops (coupling angle $\beta$). Classicalization of the random walk is obtained by stochastic phase  changes applied by a phase modulator (PM) at each discrete time step.}
 \end{figure}
where $\hat{a}^{\dag}_n$, $\hat{a}_n$ are the bosonic creation and destruction operators at lattice site $n$ and  $J$ is the nearest-neighbor hopping amplitude. Let us assume that at initial time $t=0$ a single photon is injected into the site $n=0$ of the lattice, i.e. the initial photon state is given by $| \psi(t=0) \rangle=\hat{a}^{\dag}_0 | 0 \rangle$. Under coherent evolution, at subsequent time $t$ the photon field is  in the state $ |\psi(t) \rangle=\sum_n \psi_n(t) \hat{a}_n^{\dag} |0 \rangle,$
 where $\psi_n(t)=\sum_l \mathcal{U}_{n,l}(t) \psi_l(0)=U_{n,0}$ and $\mathcal{U}=\exp(-i  H t)$ is the coherent propagation (scattering) matrix, with $H_{n,l}=J(\delta_{n,l+1}+\delta_{n,l-1})$. In a coupled waveguide system, such as in the waveguide array shown in Fig.1(a), the evolution time $t$ is replaced by the propagation distance $z$ along the waveguide lattice with the correspondence $z=ct$, where $c$ is the speed of light \cite{R31b}. The probability $P_n(t)$ that a photo-detector placed at site $n$ 'clicks' and destroys the photon at time $t$ is given by $P_n(t)= |\psi_n(t)|^2$. For coherent evolution, such a probability distribution reads explicitly $P_n(t)=J_n^2( 2 J t)$, where $J_n$ is the Bessel function of first kind and order $n$. Correspondingly, the mean photon position is  $\langle n(t) \rangle=\sum_n n P_n(t)=0$ while the variance of the probability distribution reads 
$ \langle n^2(t) \rangle=\sum_n n^2 P_n^2(t)=2J^2t^2$,
 corresponding to ballistic spreading of the photon in the lattice.  Here $\langle.. \rangle$ denotes the average due to intrinsic quantum randomness. We mention that, since we consider the dynamics of a single photon, quantum interference of more than one quantum particle, i.e. multiphoton quantum interference, is absent in the coherent walk and the distribution probability $P_n(t)$ relies only on the wave-like interferences of single-particle probability amplitudes $\psi_n$. As a consequence, it is well known that $P_n(t)$ can be obtained in a purely classical setting using intense light fields exploiting the wave nature of classical  light (see e.g. \cite{R32}). In this case the probability distribution $P_n(t)$ is obtained in a single-shot measurement from the intensity distribution of the light in the various lattice sites [bottom panels in Fig.1(a)], i.e. there is not need to perform repeated measurements.\\ 
The transition from quantum to classical behavior in the random walk is driven by decoherence \cite{R5}. For photons, decoherence can be analyzed and controlled in various ways, such as by phase damping or measurements (see e.g. \cite{R7,R12,R15,R16,R18,R20,R22,R23,R33,R34}). For example, random phase shifts can be periodically applied to the wave amplitudes $\psi_n$ at times $t_{\alpha}= \alpha \Delta t$ spaced by the interval $\Delta t $ ($\alpha=1,2,3,..$), leading to loss of coherence without energy exchange. Under dephasing dynamics,
 the mean value $\mathbb{E}[O]$ of any observable $O$ is computed as  
 \[ \mathbb{E}[O]=\overline{\langle \psi(t) | O |\psi(t) \rangle},\]
  where the overbar denotes ensemble averaging over different realizations of the random phase distributions, i.e. over all possible quantum trajectories $| \psi (t) \rangle$  evolving under a specific realization of stochastic phases. The mean value $\mathbb{E}[O]$, which comprises both intrinsic quantum uncertainty ($\langle ... \rangle$) and ensemble ($ \overline{^{}...}$) averaging, can be compactly computed in terms of the density operator $\rho(t)= \overline{| \psi(t) \rangle \langle \psi(t)|}$ via the trace relation $\mathbb{E}[O]={\rm Tr} ( \rho O)$. For a photonic random walk on a line, we can introduce the density matrix $\rho_{n,m}(t)=\overline{\psi_n^*(t) \psi_m(t)}$. Clearly, $\rho_{n,m} (t_{\alpha}^+)=0$ for $n \neq m$, where $t=t_{\alpha}^+$ is the time instant successive to the application of the random phase: basically, dephasing drives the dynamics into the classical regime (CRW), which is fully described by a discrete-time map for the photon distribution probabilities $P_n(t)= \rho_{n,n}(t)$ at the various lattice sites.
Assuming a short time interval $\Delta t \ll 1/J$ between successive stochastic phases, such that ${U}(\Delta t) =\exp(-i H \Delta t)$ can be expanded up to second order in $J \Delta t$ as ${U} \simeq 1-i H \Delta t-(1/2) H^2 \Delta t^2$,  the average occupation probabilities satisfy the Markovian master equation (see for instance \cite{R33,R34} and the Supplemental document)
\begin{equation}
\frac{dP_n}{dt}=J_e \left\{ -2 P_n(t)+P_{n+1}(t)+P_{n-1}(t) \right\}
\end{equation}
which is characteristic of a CRW on the line. In the above equation, $J_e=J^2 \Delta t$ is the effective incoherent hopping rate of the photon between adjacent sites of the lattice. The same classical master equation can be obtained in the framework of the Lindblad master equation for the density operator in the strong dephasing regime \cite{R5,R11,R20,R27,R34,R35}. The solution to Eq.(2) with the initial condition $P_n(0)=\delta_{n,0}$ reads $P_n(t)=I_n(2 J_et) \exp(-2 J_et) $, where $I_n$ is the modified Bessel function of order $n$. In the classical limit of photonic random walk, the mean position and variance of the walker are given by $\mathbb{E}[n]=\sum_{n} nP_n(t)=0$ and $\mathbb{E}[n^2]=\sum_n n^2 P_n(t)=2J_et,$
 corresponding to a diffusive spreading of the photon on the lattice. The transition from quantum to classical random walk is illustrated in Figs. 2(a) and 2(b).\\
\\
{\it Subdiffusive dynamics in the photonic random walk.}  An intriguing aspect of the photonic random walk under dephasing dynamics is the emergence of subdiffusive  dynamics of the photon wave packet, observable when disentangling quantum and ensemble averaging. Although such disentanglement is not feasible when the lattice is excited by a single photon, it becomes quite straightforward when the lattice is probed with classical light states carrying a large number of photons. Instead of considering the mean $\mathbb{E}[n^2]$, let us examine the square of the distance 
$d$ of the center of mass of the light intensity distribution $|\psi_n(t)|^2$ from the initially excited site $n=0$, given by
\begin{equation}
{d^2}=\left(  \sum_n n | \psi_n(t)|^2 \right)^2 = \langle n \rangle ^2 .
\end{equation}
 The square distance $d^2$ is readily measured when the system is excited by a classical light field carrying a large number of photons, since in this case there is no need to make any statistical average related to quantum randomness and the classical light intensity distribution $|\psi_n(t)|^2$ already embodies the quantum statistical average.
 Clearly, the distance $d$ depends on the specific realization of the stochastic phases, i.e. on the specific trajectory $|\psi(t) \rangle$ evolving according to a given sequence of applied random phases. 
   We then compute the ensemble average of $d^2$, i.e. the quantity
 \begin{equation}
 \overline{d^2}= \overline{\langle n \rangle^2}
 \end{equation}
which is obtained by repeated classical light propagation measurements in the system for different realizations of stochastic phases. The main and key point is that, since
$\mathbb{E}[n^2]=\overline{\langle n^2 \rangle } \neq \overline{\langle n \rangle ^2}$, the asymptotic behavior of $\overline{d^2}$ versus time $t$ is no longer linear, i.e. the motion of the wave packet center of mass is not diffusive.  This would be true if and only if, at any time $t$, $\psi_n(t)$ is localized with certainty at a given site, i.e. if the walker would be a classical particle, in such a way that $ \overline{\langle n \rangle ^2}=\overline{\langle n^2 \rangle }$.  While 
$\mathbb{E} [ n^2 ] = \overline{\langle n^2 \rangle}$ can be computed using the density matrix $\rho_{n,m}= \overline{\psi_n^*(t) \psi_m(t)}$ according to the trace rule $\mathbb{E}[n^2]=\sum_n n^2 \rho_{n,n}$, the mean square distance $\overline{d^2}$ cannot be calculated from the knowledge of the density matrix solely since it involves second-order correlations of the stochastically-evolving amplitudes $\psi_n(t)$ \cite{R25,R27,R36}. Namely one has 
\begin{equation}
\overline{d^2}= \overline{\left( \sum_n n |\psi_n(t)|^2 \right)^2}=\sum_{n,m} n m C_{n,m}(t)
\end{equation}
where
\begin{equation}
C_{n,m}(t) \equiv \overline{| \psi_n(t)|^2 | \psi_m(t)|^2}
\end{equation}
is the ensemble-averaged correlation of the occupation probabilities. Interestingly, it can be shown that  from a formal viewpoint the ensemble-averaged correlation $C_{n,m}(t)$ describes a classical random walk on a two dimensional square lattice with symmetric vertical/horizontal hopping probability $J_e=J^2 \Delta t$, except for the sites on the three diagonals  $n=m, m \pm 1$ where the hopping  probabilities are asymmetric. Details are provided in the Supplemental document. 
A central result, also shown in the Supplemental document and consistent with previous studies \cite{R25,R26,R36}, is that $\overline{d^2}$ asymptotically increases with time $t$ as $\sim \sqrt{t}$, indicating that the center of mass of the classical light wave packet  undergoes a {\em subdiffusive} motion, i.e. very distinct than both the diffusive motion of a classical walker and the ballistic motion of a quantum walker.  An example of subdiffusive dynamics  is shown in Fig.2(c). The main reason thereof is that, by probing the system with an intense (classical) light field and disentangling quantum averaging from ensemble averaging, the mean square distance of the wave packet center of mass versus time $t$ is dictated by second-order correlations $C_{n,m}(t)$ of the stochastic wave functions, rather than by the first-order correlations as in the density matrix description of classicalization of the photonic quantum walk. Basically, for each specific realization of stochastic phases the center of mass of the classical wave packet with intensity distribution $|\psi_n(t)|^2$ 
follows an irregular path, and rather generally at any time $t$ it is not found in its initial position $n=0$. The square distance $d^2$ of the wave packet center of mass, when averaged over all possible different paths, does not increase linearly with time, which would be expected for a classical walker that at any time $t$ is with certainty in a given site of the lattice, rather than distributed along several lattice sites with the distribution $|\psi_n(t)|^2$. In other words, the classical light distribution retains some memory of the quantum indeterminacy (i.e. state superposition) of the walker, which is reflected in the final subdiffusive motion of its center of mass. Conversely, when the system is excited by a single photon, the statistical average of photon position for different experimental runs cannot disentangle quantum and ensemble randomness, which results in a fully classicalization of the random walk and diffusive spreading dynamics.\\

      \begin{figure}[h]
 \centering
    \includegraphics[width=0.49\textwidth]{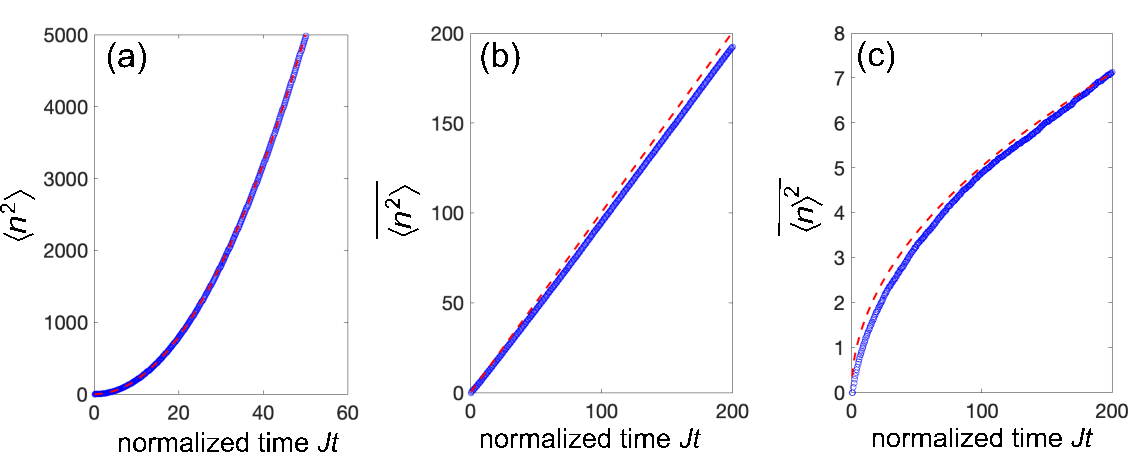}
   \caption{ \small Different spreading dynamics in the quantum-to-classical photonic random walk transition.  (a) Behavior of the mean variance $\langle n^2 \rangle$ versus propagation time $t$ in the coherent quantum walk regime (single photon excitation), corresponding to ballistic spreading $\langle n^2 \rangle=2 J^2t^2$. (b) Same as (a), but with dephasing (diffusive spreading). (c) Evolution of the mean square distance  $\overline{\langle n \rangle^2}$ in the photonic random walk with dephasing when initial excitation is an intense classical light field. Parameter values are $J=1$ (coherent hopping rate) and $\Delta t=0.5$. The bold blue curves in (b) and (c) are obtained after ensemble averaging over 10000 realizations of stochastic phases, applied every $\Delta t$ time step. The dashed red line in (b) is the theoretically-predicted curve $ \overline{\langle n^2 \rangle}=2J_et$, with $J_e=J^2 \Delta t$. The dashed red line in (c) is the fitting curve $\overline{\langle n \rangle^2} \simeq 0.72 \sqrt{J_et}$.  }
\end{figure}

 {\it Subdiffusive dynamics in synthetic temporal lattices}.  To illustrate the subdiffusive dynamics  in an experimentally-accessible photonic setting, let us consider light dynamics in synthetic temporal lattices based on optical pulse circulation in coupled fiber loops \cite{R28,R29,R30,R31}. The system  consists of two fiber loops of slightly different lengths $L \pm \Delta L$, that are connected by a variable directional coupler with a coupling angle $\beta$, as schematically shown in Fig.1(b). A phase  modulator is placed in one of the two loops, which introduces stochastic phase changes thus emulating decoherence \cite{R16,R34,R37}.  When a single optical pulse is injected into one loop, it will periodically split and interfere, evolving into a pulse train and realizing a synthetic lattice via time multiplexing \cite{R28,R29,R30,R31}. By discretizing the physical time as $t=t_n^{(m)}= mT +n\Delta T$, where $T= L/c$ is mean travel time and $\Delta T=\Delta L/c \ll T$ is the travel-time difference of light pulses in two loops, light dynamics is described by the set of discrete-time equations (see e.g. \cite{R28,R29,R30,R31,R33,R34,R37})
 \begin{eqnarray}
u_n^{(m+1)} & = & \left( \cos \beta u_{n+1}^{(m)}+i \sin \beta w_{n}^{(m)} \right) \exp(i \phi_n^{(m)}) \;\;\;\;\;\; \\
w_n^{(m+1)} & = & i \sin \beta u_{n}^{(m)}+ \cos \beta w_{n-1}^{(m)}.
\end{eqnarray}
In the above equations, $u_n^{(m)}$ and $w_n^{(m)}$ are the pulse amplitudes at discrete time step $m$ and lattice site (unit cell) $n$ in the two fiber loops and $\phi_n^{(m)}$ are uncorrelated stochastic phases with uniform distribution in the range $(-\pi, \pi)$.  A correspondence between the discrete-time random walk, defined by  Eqs.(7,8), and the continuous-time random walk, investigated in previous sections and described by the coherent Hamiltonian (1), can be formally established in the limit $\beta \rightarrow \pi/2$ (see for instance \cite{R34}); in this limiting case Eqs.(7) and (8) describe a  continuous-time random walk with a coherent hopping rate $J=(1/2) \cos \beta$, a dephasing time interval $\Delta t=2$, corresponding to an incoherent hooping rate $J_e=J^2 \Delta t= (1/2) \cos^2 \beta$, and $t=m$ for the time variable. A typical evolution dynamics, for a coupling angle $\beta= 0.8 \times \pi/2$, is shown in Fig.3. Under coherent dynamics [Fig.3(a)], the numerically-computed behavior of the variance $\langle n^2 \rangle=\sum_n n^2(|u_n^{(m)}|^2+ |w_n^{(m)}|^2)$ shows a parabolic increase with time step $m$, corresponding to ballistic spreading. Conversely, after phase randomization at each time step  the ensemble average variance
$\overline{\langle n^2 \rangle}=\sum_n n^2 \overline{(|u_n^{(m)}|^2+ |w_n^{(m)}|^2)}$ shows a linear increase with $m$ [diffusive spreading, Fig.3(b)], while  $\overline{ \langle n \rangle^2}$ displays subdiffusive dynamics [Fig.3(c)], according to the theoretical analysis.

  \begin{figure}[h]
 \centering
    \includegraphics[width=0.49\textwidth]{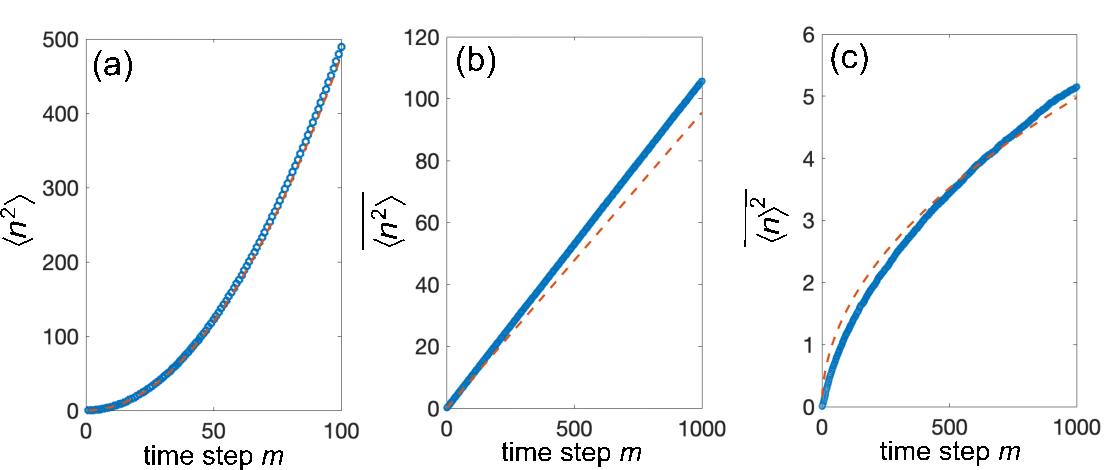}
   \caption{ \small Different spreading dynamics in the quantum-to-classical photonic random walk transition realized by the synthetic temporal lattice in the coupled fiber loop system of Fig.1(a). Coupling angle used in the numerical simulations is  $\beta= 0.8 \times \pi/2$; initial condition is $u_n^{(0)}=\delta_{n,0}$ and $w_n^{(0)}=0$, corresponding to single pulse injection in one loop.  Panels (a), (b) and (c) are the same as in Fig.2.  In (b) and (c) the bold blue curves are obtained after ensemble averaging over 10000 realizations of stochastic phases, whereas the dashed red lines are the theoretical curves $\overline{\langle n^2 \rangle}=2J_e m$ and $\overline{\langle n \rangle ^2} \simeq 0.72 \sqrt{J_e m}$ obtained by assuming an effective incoherent hopping amplitude $J_e=(1/2) \cos^2 \beta$.}
\end{figure}

{\em Conclusion.}
The quantum-to-classical transition of a photon's random walk in a tight-binding lattice under decoherence is commonly indicated by a shift from ballistic to diffusive spreading under single photon excitation \cite{R16}. However, when intrinsic quantum randomness and ensemble randomness are disentangled, a subdiffusive dynamics for the light packet's center of mass can be observed, which is possible by simply exploiting the wave nature of intense classical light fields. Here we have shown that such a striking anomalous behavior can be explained from an effective random walk with defects on a two-dimensional lattice for the correlations of occupation probabilities, and suggested discrete-time photonic random walks in synthetic temporal lattices as experimentally accessible testbed of predicted phenomena.\\
\\ 
\noindent
{\bf Disclosures}. The author declares no conflicts of interest.\\
\\
{\bf Data availability}. No data were generated or analyzed in the presented research.\\
\\
{\bf Funding}. Agencia Estatal de Investigacion (MDM-2017-0711).\\
\\
{\bf Supplemental document}. See Supplement 1 for supporting content.

\newpage

\onecolumn

{\bf Supplemental document}

\section{Quantum to classical transition in the continuous-time photonic random walk}

The transition from quantum to classical behavior in a random walk is driven by decoherence \cite{R5}. For photons, decoherence can be analyzed and controlled in various ways, such as by phase damping or measurements (see e.g. \cite{R7,R12,R15,R16,R18,R20,R22,R23,R33,R34}). The evolution of quantum walks with decoherence is non-unitary, thus the density operator $\rho$ is needed for describing mixed states. Usually, decoherence is embodied in a master equation for the density operator of Lindblad form, with local dissipators describing pure dephasing at a rate $\gamma$. In the strong dephasing limit $\gamma \gg J$, the quantum walk is equivalent to a classical random walk (see for instance \cite{R34}).
In a photonic setting, dephasing can be introduced by random phase shifts, periodically applied to the wave amplitudes $\psi_n$ at times $t_{\alpha}= \alpha \Delta t$ spaced by the interval $\Delta t $ ($\alpha=1,2,3,..$), leading to loss of coherence without energy exchange. The periodic phase randomization of the wave amplitudes $\psi_n$ at time intervals $\Delta t \ll 1/J$ drives the photonic random walk into the fully classical regime. In fact let us introduce the density matrix $\rho_{n,m}(t)=\overline{\psi_n^*(t) \psi_m(t)}$, where the overline indicates ensemble average.
At any time $t=t_{\alpha}^+$ one clearly has $\rho_{n,m}=0$ for $n \neq m$, while $\rho_{n,m}$ remains small at any other time $t$. Hence, dephasing drives the dynamics into the classical regime, which is fully described by  the diagonal density matrix elements  $\rho_{n,n}(t)=P_n(t)$, corresponding to the photon distribution probabilities at the various lattice sites. To derive the evolution equations for the density matrix elements $\rho_{n,n}(t)$, let us calculate
\begin{equation}
\rho_{n,n}(t+\Delta t)=\overline {\psi_n(t+\Delta t) \psi_n^*(t+\Delta t)}= \sum_{l, \sigma} U^*_{n,l}(\Delta t) U_{n,\sigma}(\Delta t) \overline{ \psi_l^*(t) \psi_{\sigma}(t)}=\sum_{l} |U_{n,l}(\Delta t)|^2 \rho_{l,l} (t)
\end{equation} 
given that $\overline{ \psi_l^*(t) \psi_{\sigma}(t)}=\rho_{l, \sigma} (t) =\delta_{l,\sigma} \rho_{l,l}(t)$. In the above equation, $U(\Delta t) = \exp(-i H \Delta t)$ is the coherent propagator for a time interval $\Delta t$ and $H$ is the single-particle tight-binding matrix Hamiltonian with elements 
\begin{equation}
H_{n,m}=J( \delta_{n,m+1}+\delta_{n,m-1}). 
\end{equation}
 Since $ J \Delta t \ll1$, we can expand $U(\Delta t)=\exp(-i \Delta t H)$ in power series of $J \Delta t$. Up to second order, one has
\begin{equation}
U_{n,l}(\Delta t) \simeq \delta_{n,l}-i H_{n,l} \Delta t-\frac{1}{2} \Delta t^2 (H^2)_{n,l}
\end{equation}
so that 
\begin{equation}
|U_{n,l}(\Delta t)|^2 \simeq \delta_{n,l}+\Delta t^2 \left\{  H^2_{n,l}-\delta_{n,l} (H^2)_{n,n} \right\}.
\end{equation}
Taking into account that
\begin{equation}
(H^2)_{n,n}=\sum_{\sigma} H_{n, \sigma} H_{\sigma,n}. 
\end{equation}
substitution of Eqs.(S2) and (S5) into Eq.(S4) yields
\begin{equation}
|U_{n,l}(\Delta t)|^2 \simeq \delta_{n,l}( 1-2J^2 \Delta t^2) + J^2 \Delta t^2 ( \delta_{n,l+1}+\delta_{n,l-1} ).
\end{equation}
From Eqs.(S1) and (S6) one then obtains
\begin{equation}
\rho_{n,n}(t+\Delta t) \simeq \rho_{n,n}(t)-2 J^2 \Delta t^2 \rho_{n,n}(t)+J^2 \Delta t^2 (\rho_{n+1,n+1}(t)+\rho_{n-1,n-1}(t) ),
\end{equation}
i.e.
\begin{equation}
\frac{d \rho_{n,n}}{dt}=-2J_e \rho_{n,n}+J_e \rho_{n+1,n+1}+J_e \rho_{n-1,n-1}
\end{equation}
which is the classical master equation (2) given in the main text, where $J_e=J^2 \Delta t$ is the effective incoherent hopping rate.

\section{Master equation for the correlation function and subdiffusive dynamics}
As discussed in the main text, the ensemble-averaged square distance $\overline{\langle n \rangle^2 }$ of the wave packet center of mass $\langle n \rangle (t)$, for the initially-excited site $n=0$, is given by
\begin{equation}
\overline{\langle n \rangle^2 }(t)= \overline{\left( \sum_n n |\psi_n(t) |^2 \right)^2}= \sum_{n,m} n m C_{n,m}(t)
\end{equation}
where 
\begin{equation}
C_{n,m}(t) = \overline{|\psi_n(t)|^2 |\psi_m(t)|^2}
\end{equation}

  \begin{figure}[ht]
 \centering
    \includegraphics[width=0.7\textwidth]{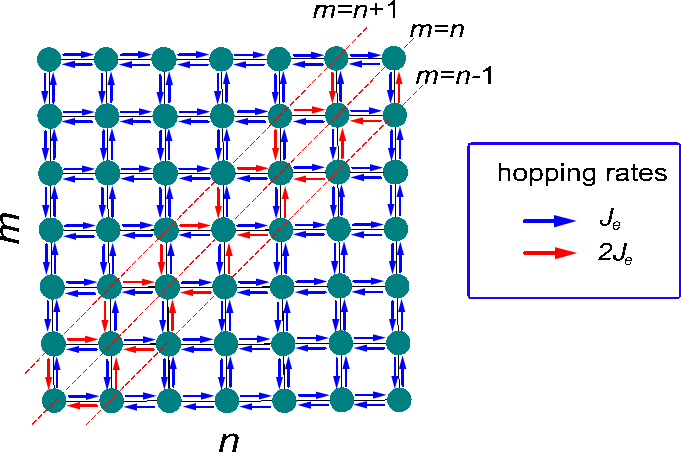}
  \caption{ Schematic of a two-dimensional random walk of a classical particle on a square lattice with asymmetric hopping rates along the diagonals $m=n \pm 1$. The strength of hopping rates is indicated by the blue/red arrows, with $J_e=J^2 \Delta t$. The master equation describing the hopping dynamics of a classical particle on the lattice is given by Eq.(S14), where $C_{n,m}(t)$ is the probability to find the particle at lattice site $(n,m)$ at time $t$, starting from the site $(0,0)$.}
\end{figure}
is the ensemble-averaged correlation of the occupation probabilities. In this section we show that \\ 
(i) The evolution equation of the correlation $C_{n,m}(t)$ formally describes a classical two-dimensional random walk on a square lattice with symmetric vertical/horizontal hopping probability $J_e=J^2 \Delta t$, except for the sites on the three diagonals  $n=m, m \pm 1$ where the hopping  probabilities are asymmetric, as schematically shown in Fig.S1.\\
 (ii) The spreading law of $\overline{\langle n \rangle^2 }(t)$ must be subdiffusive, namely $\overline{\langle n \rangle^2 } \sim \sqrt{t}$.\\
To prove the above statements, let us proceed as in the previous Sec.1 and calculate $C_{n,m}(t+\Delta t)$, which reads explicitly
\begin{equation}
C_{n,m}(t+\Delta t)=\sum_{l,l',\sigma,\sigma'}U_{n,l} (\Delta t) U^*_{n,l'}  (\Delta t) U_{m, \sigma}  (\Delta t) U^*_{m,\sigma '}  (\Delta t) \overline{\psi_l(t) \psi_{l'}^*(t) \psi_{\sigma}(t) \psi_{\sigma'}^*(t)}.
\end{equation}
Clearly, owing to phase randomization the ensemble average $\overline{\psi_l(t) \psi_{l'}^*(t) \psi_{\sigma}(t) \psi_{\sigma'}^*(t)}$ vanishes unless (i) $l=l'$ and $\sigma= \sigma'$, or (ii) $l= \sigma'$ and $\sigma=l'$ (with $l \neq \sigma$). Hence one has
\begin{equation}
C_{n,m}(t+\Delta t)=\sum_{l,\sigma} |U_{n,l} (\Delta t)|^2 |U_{m,\sigma} (\Delta t)|^2 C_{l,\sigma}(t)+\sum_{l \neq \sigma} U_{n,l}(\Delta t) U_{m,\sigma}(\Delta t) U_{n,\sigma}^*(\Delta t) U_{m,l}^*(\Delta t) C_{l,\sigma}(t).
\end{equation}
The various terms entering on the right hand side of Eq.(S12) can be calculated, after some lengthy but straightforward calculations, up to second order in the small parameter $J \Delta t$ by letting $U(\Delta t)=\exp(-iH \Delta t) \simeq 1-iH \Delta t-(1/2) \Delta t^2 H^2$. This yields
\begin{eqnarray}
C_{n,m}(t+\Delta t) & = & C_{n,m}(t)+J^2 \Delta t^2 \left\{ C_{n,m+1}(t)+C_{n,m-1}(t)+C_{n+1,m}(t)+C_{n-1,m}(t) -4 C_{n,m}(t) \right\} \;\;\;\; \nonumber \\
& + & J^2\Delta t^2 \delta_{n,m} \left\{ C_{n,n-1}(t)+C_{n,n+1}(t)+C_{n+1,n}(t)+C_{n-1,n}(t)  \right\} \\
& - & 2J^2 \Delta t^2 (\delta_{n,m+1}+\delta_{n,m-1}) C_{n,m}(t). \nonumber
\end{eqnarray}
In the $\Delta t J \rightarrow 0$ limit, one then obtains the following classical master equation for the correlation $C_{n,m}(t)$
\begin{eqnarray}
\frac{d C_{n,m}}{dt} & = & J_e \left\{ C_{n,m+1}+C_{n,m-1}+C_{n+1,m}+C_{n-1,m} -4 C_{n,m}\right\}  \\
& + & J_e \delta_{n,m} \left\{ C_{n,n-1}+C_{n,n+1}+C_{n+1,n}+C_{n-1,n}  \right\} - 2J_e (\delta_{n,m+1}+\delta_{n,m-1}) C_{n,m}.  \nonumber
\end{eqnarray}
where we have set $J_e=J^2 \Delta t$. From a formal viewpoint the above master equation describes a classical random walk on a two-dimensional square lattice with symmetric vertical/horizontal hopping probability $J_e=J^2 \Delta t$, except for the sites on the three diagonals  $n=m, m \pm 1$ where the hopping  probabilities are asymmetric, as schematically shown in Fig.S1.
The spreading dynamics of a classical particle on such a square lattice, initially placed at the site $n=m=0$ and corresponding to the initial condition $C_{n,m}(0)=\delta_{n,0}\delta_{m,0}$, basically determines the temporal behavior of  $\overline{\langle n \rangle^2 }(t)$, according to Eq.(S9).  Typical probability distributions $C_{n,m}(t)$ at subsequent times, obtained by numerically solving  Eq.(S14), are shown in Fig.S2. As it can be seen, the classical particle spreads around the initial position $n=m=0$ isotropically, with $C_{n,m}(t)=C_{m,n}(t)=C_{-n,-m}(t)$ like in an homogeneous two-dimensional classical random walk on a square lattice, except that a  higher probability is observed to find the particle along the main diagonal $n=m$. Such isotropy breaking  arises from the asymmetry of hopping rates among the sites on the diagonals $m=n,n \pm 1$, which favors the particle to remain on the main diagonal $n=m$. In the absence of the asymmetry, i.e. by neglecting the terms in the second raw of Eq.(S14), the particle would describe a standard two-dimensional random walk and the  exact solution to Eq.(S14) can be expressed in terms of modified Bessel functions $I_n$ as 
$C_{n,m}(t)= \exp(-4J_et) I_n(2J_et) I_m(2J_et)$, which in the long time limit $J_e t \gg 1$ is well approximated by the isotropic two-dimensional Gaussian distribution
\begin{equation}
C_{n,m}(t) \simeq \frac{1}{4 \pi J_e t} \exp \left(  -\frac{n^2+m^2}{4J_e t} \right).
\end{equation}
  \begin{figure}[ht]
 \centering
    \includegraphics[width=0.98\textwidth]{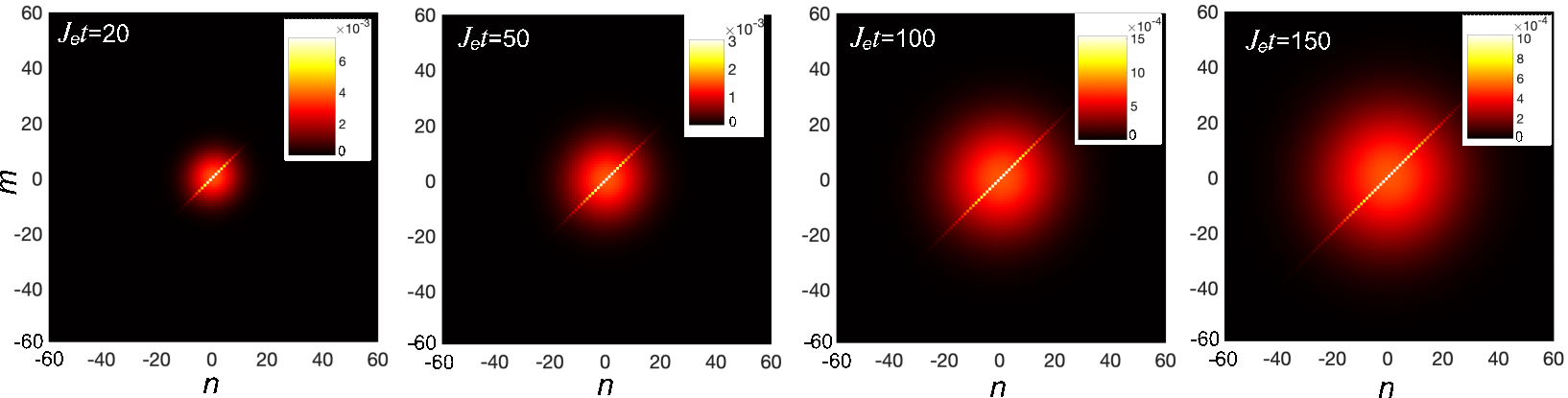}
  \caption{  Two-dimensional random walk of a classical particle in the square lattice of Fig.S1 with asymmetric hopping rates along the diagonals $m=n \pm 1$. The particle is initially placed at site $n=m=0$. The four panels show on a pseudocolor map the  numerically-computed evolution of $C_{n,m}(t)$ for a few increasing values of $J_et$.}
\end{figure}
When the asymmetry in the hopping rates along the diagonals $n=m \pm 1$ are considered, an exact analytical solution to Eq.(S14) is challenging to obtain, and one has to resort to numerical simulations. However, in order to determine the asymptotic temporal behavior of $\overline{\langle n \rangle^2}(t)$, it is sufficient to provide approximate form of the asymptotic solution. To this aim, from the numerical simulations and given the form of Eq.(S14), its asymptotic solution  for $J_e t \gg 1$  can be roughly estimated by the following anisotropic distribution, which generalizes Eq.(S15) to account for asymmetric hopping rates
\begin{equation}
C_{n,m}(t) \simeq \left\{
\begin{array}{cc}
\frac{\beta}{4 \pi J_e t} \exp \left(  -\frac{n^2+m^2}{4J_e t} \right) & n \neq m \\
\frac{\beta}{2 \pi J_e t} \exp \left(  -\frac{n^2}{2J_e t} \right)  & n=m
\end{array}
\right.
\end{equation}
In the above equation, the multiplicative constant $\beta$ should be determined by probability conservation $\sum_{n,m}C_{n,m}(t)=1$ and reads
\begin{equation}
\beta=\frac{1}{1+\frac{2}{2 \sqrt{2 \pi J_e t}}} \simeq 1
\end{equation} 
for $J_e t \gg 1$. Basically, the solution (S16) is a smooth function of $n$ and $m$, except for $n=m$ where it is discontinuous as the main diagonal $n=m$ is crossed; the value of the function on the main diagonal $n=m$ is twice than in the neighbor sites, in order to match the defective terms in Eq.(S14) .  We checked that the approximate asymptotic behavior of $C_{n,m}(t)$ for $J_et \gg 1$, as given by Eq.(S16), provides a good approximation to the exact numerically-computed solution, shown in Fig.S2, except than in the tails of the distribution. Substitution of Eq.(S16) into Eq.(S9) and after making the summation it readily follows that 
\begin{equation}
\overline{\langle n \rangle^2}(t)=\alpha \sqrt{J_e t}
\end{equation}
where the multiplication constant is given by $\alpha=(2 \sqrt{2 \pi})^{-1}$. This result indicates that the mean square distance of the wave packet increases in time subdiffusively. The numerical results confirm such a subdiffusive behavior, with a corrected multiplication constant $\alpha \simeq 0.72$, as shown in Fig.S3. 

  \begin{figure}[ht]
 \centering
    \includegraphics[width=0.8\textwidth]{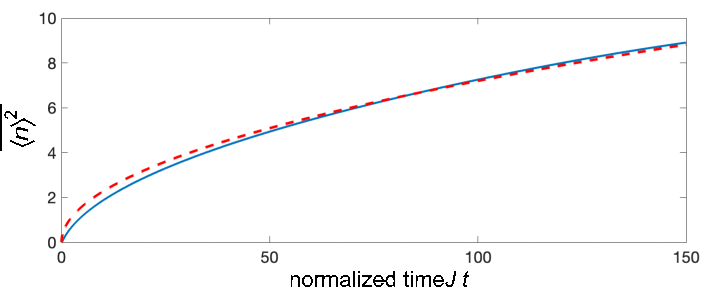}
  \caption{  Numerically-computed temporal evolution of $\overline{\langle n \rangle^2}(t)= \sum_{n,m} n m C_{n,m}(t)$ versus normalized time $J_et$ (solid blue curve), and fitting curve $\overline{\langle n \rangle^2}=\alpha \sqrt{J_et}$ with $\alpha=0.72$ (dashed red curve).}
\end{figure}

\end{document}